\documentclass[12pt]{article}

\usepackage{newtxtext,newtxmath}
\usepackage{comment}
\usepackage{placeins}

\usepackage{graphicx}

\usepackage[letterpaper,margin=1in]{geometry}
\usepackage{comment}

\renewenvironment{abstract}
	{\quotation}
	{\endquotation}

\date{}

\makeatletter
\renewcommand{\fnum@figure}{\textbf{Figure \thefigure}}
\renewcommand{\fnum@table}{\textbf{Table \thetable}}
\makeatother

\usepackage{scicite}

\usepackage{caption}
\usepackage{hyperref}

\def\scititle{
	Observation of Super-ballistic Brownian Motion in Liquid
}
\title{\bfseries \boldmath \scititle}

\author{
	Jason Boynewicz$^{\ast\dagger}$,
	Michael C. Thumann$^{\dagger}$,
	Mark G. Raizen$^{}$\and
	\small$^{}$Department of Physics, The University of Texas at Austin, Austin, Texas 78712, USA\and
	\small$^\ast$Corresponding author. Email: jrb8338@my.utexas.edu\and
	\small$^\dagger$These authors contributed equally to this work.
}

\begin{document} 

\maketitle

\begin{abstract} \bfseries \boldmath
Brownian motion is a foundational physical process characterized by a mean squared displacement that scales linearly in time in thermal equilibrium, known as diffusion. At short times, the mean squared displacement becomes ballistic, scaling as $t^2$. This effect was predicted by Einstein in 1907 and recently observed experimentally. We report that this picture is only true on average; by conditioning specific initial velocities, we predict theoretically and confirm by experiment that the mean squared displacement becomes super-ballistic, with a power scaling law of $t^{5/2}$. This result is due to the colored noise of incompressible fluids, resulting in a non-zero first moment for the thermal force when conditioned on non-zero initial velocities. These results are a step towards the unraveling of nonequilibrium dynamics of fluids.
\end{abstract}

\noindent
\section{Introduction}
Brownian motion lies at the intersection of fluid mechanics, molecular dynamics, and statistical physics. Historically, the work of Einstein, Smoluchowski, Sutherland, and Perrin proved the atomistic hypothesis of matter \cite{einstein_uber_1905, smoluchowski_kinetic_1906, sutherland_lxxv_1905, perrin_mouvement_1909}. The key insight of these early works was the role of a stochastic force generating the thermal motion \cite{langevin_sur_1908}. Since then, models that describe these stochastic forces have evolved to account for inertial, viscoelastic, and even compressible effects \cite{vladimirsky_hydrodynamical_1945, hinch_application_1975, berne_calculation_1966, schram_theory_1998}. While these models have proven very successful experimentally, experiments have focused on equilibrium correlation functions or overdamped dynamics. With experimental access to the velocity of a spherical Brownian particle in a Newtonian fluid, we examine the stochastic forcing characteristic to hydrodynamic non-Markovian Brownian motion. By preferentially selecting moments from equilibrium in which the particle velocity is close to zero, we diminish the deterministic forcing terms, allowing fluctuation to temporarily dominate dissipation. We treat this study as a lens into the fluctuation-dissipation correspondence, and view conditioning as a study on nonequilibrium dynamics for an inertial Brownian particle in a Newtonian fluid. \par

 The invention of optical tweezers by Arthur Ashkin has allowed the trapping and tracking of mesoscopic particles subject to substantial thermal fluctuations \cite{ashkin_observation_1986}. Advances in back focal plane interferometry \cite{denk_optical_1990, svoboda_direct_1993} have permitted the tracking of a Brownian microsphere well below its momentum relaxation time \cite{kheifets_observation_2014, madsen_ultrafast_2021}. These advances have allowed measurement of the transition to ballistic motion \cite{huang_direct_2011}, the Maxwell-Boltzmann distribution in both gas and liquid \cite{li_measurement_2010, mo_testing_2015}, and the effect of hydrodynamic memory on Brownian motion \cite{franosch_resonances_2011, hammond_direct_2017}. The result has been experimental confirmation of the equilibrium correlation functions and mean-squared displacement (MSD) for a Brownian particle in an incompressible fluid. \par In addition to these equilibrium short-time studies, researchers have used mesoscopic systems in fluids to test nonequilibrium results and stochastic thermodynamics \cite{liphardt_equilibrium_2002, gomez-solano_energy_2024, ciliberto_experiments_2017}. Nevertheless, these mesoscopic experiments in dense fluids occur exclusively in the overdamped regime, where there is no access to the velocity degree of freedom.
 Levitated nanoparticle experiments have been successful in probing inertial nonequilibrium phenomena in Markovian environments \cite{gieseler_dynamic_2014,hoang_experimental_2018}. In a gas, the small ratio between the density of the fluid $\rho_f$ and the density of the Brownian particle $\rho$ leads to an apparent white noise forcing term and the familiar Langevin equation. For a spherical particle of mass $m$ and radius $a$ suspended in a fluid with viscosity $\eta$ and held with trap strength $K$, the one-dimensional dynamics are governed by
 
\begin{align}
 &m\ddot{x}(t) = -\gamma \dot{x}(t) - K x(t) + R(t)\\
&\gamma = 6\pi a \eta, \quad \tau_p = \frac{m}{\gamma} \\
&\langle R(t)R(t') \rangle = 2k_BT\gamma\delta(t - t')\, .
\end{align}

When the two densities become more comparable in a liquid, the inertia of the fluid generates an extra contribution to the force exerted on the particle by the fluid, which involves the history of the particle. By the fluctuation-dissipation theorem \cite{kubo_fluctuation-dissipation_1966}, the stochastic forcing term gains a colored component \cite{hinch_application_1975}. Under the assumption of no-slip boundary conditions, these considerations yield the hydrodynamic generalized Langevin equation (GLE) of the form

\begin{align}
        & M \ddot{x}(t) = -\gamma \dot{x}(t) -\gamma\sqrt{\frac{\tau_f}{\pi}}\int_{-\infty}^t\frac{\ddot{x}(\tau)}{\sqrt{t - \tau}}d\tau -Kx(t) + R(t) \label{h_GLE}\\
    & \langle R(t)R(t') \rangle = 2 \gamma k_BT (\delta(t-t') - \frac{1}{4}\sqrt{\frac{\tau_f}{\pi}}|t - t'|^{-3/2}) \label{equilibrium_thermal_corr}\\
    & \tau_f = \frac{\rho_f a^2}{\eta},\quad M = m +\frac{2}{3}\pi a^3\rho_f \, .
\end{align}
The Basset-Boussinesq force and colored hydrodynamic memory can be seen in the transition region between diffusive and ballistic motion \cite{franosch_resonances_2011, kheifets_observation_2014} as well as the algebraic decay of the velocity autocorrelation function (VACF) \cite{alder_velocity_1967, alder_decay_1970, zwanzig_hydrodynamic_1970}. While the transition between diffusive and ballistic regimes is altered by the hydrodynamic memory and colored noise, the free-particle asymptotic expressions in equilibrium are equivalent to the forcing case of white noise except for the addition of the added mass. With the included harmonic confinement, both MSDs tend to a stationary constant value.  \par
The correlation functions are built by averaging every trajectory measured from the experimental time trace of the particle. Different correlation functions can be built by averaging only specific trajectories with specific initial positions and velocities. For example, by conditioning the particle to begin at rest and in the center of the trap, a super-ballistic $t^3$ scaling of the MSD was shown for a Brownian particle trapped in air \cite{duplat_superdiffusive_2013}. This $t^3$ dependence can be understood as the result of acceleration from the delta-correlated stochastic force acting on the particle. At short times, the damping force and harmonic trap force are expected to be small since the particle starts at rest and in the center of the trap. Thus, for short times, the motion of the particle is governed by the differential equation
\begin{equation}
    m\ddot{x} \approx R(t) \, ,
\end{equation}
which implies the particle undergoes a free inertial process for some short time \cite{masoliver_free_1995}. Under the assumption of white noise, solving this differential equation gives a MSD of
\begin{equation}
    \operatorname{MSD}[t] \approx \frac{2}{3}\frac{k_B T}{m \tau_p}t^3 \, ,
\end{equation}
which matches the theory and experimental results shown in \cite{duplat_superdiffusive_2013}. The same differential equation can be solved with the hydrodynamic colored-noise correlation function. Doing so yields a leading-order approximate MSD of
\begin{equation}
    \operatorname{MSD}[t] \approx \frac{2}{3}\frac{k_B T}{M \tau_p} \frac{12}{5}\sqrt{\frac{\tau_f}{\pi}}t^{5/2} \, .
\end{equation}
Then, the presence of the colored noise term would alter the asymptotic form of the MSD in comparison to the white noise case. A more complete description can be obtained by expanding the VACF for a particle initialized with zero velocity, found in \cite{clercx_brownian_1992}. Expanding to second order gives
\begin{align}
    & \operatorname{MSD}[t] \approx \frac{2}{3}\frac{k_B T}{M \tau_p} (\frac{12}{5}\sqrt{\frac{\tau_f}{\pi}}t^{5/2} + \beta t^3) \label{two_term_expansion} \\
    & \beta = 1 - (1 + \frac{8}{3\pi})\frac{\tau_f}{\tau_p} \, .
\end{align}

It is worth noting that the above expression is close to the results derived for the fractional Langevin equation (FLE) for less singular memory kernels 
except with $\beta = 1$ \cite{tateishi_different_2012, sandev_langevin_2014}. For physically relevant viscoelastic models, the short time scaling is still set by $t^3$. It is only when considering the fluid inertial contribution that the asymptotic power law scaling is changed. Furthermore, in the limit of small $\rho_f$, we recover the same $t^3$ scaling demonstrated for a Brownian particle in air. Thus, by conditioning the initial velocity of the Brownian particle to be zero, the short-time behavior of the particle is determined by the correlation of the thermal noise and an asymptotic $t^{5/2}$ super-ballistic scaling of the MSD is expected. Furthermore, a similar $t^{5/2}$ term appears as a correction factor to the MSD of a Brownian particle in a shear flow \cite{clercx_brownian_1992}. This correspondence makes sense, as the shear flow carries the free inertial effects into the long time limit \cite{katayama_brownian_1996}. In an analogous way, we expect that the removal of the obscuring ballistic motion of the initial velocity should reveal the color of the thermal force on the short time MSD.

\section{Results}

\begin{figure}
\centering
\includegraphics[width=1\textwidth]{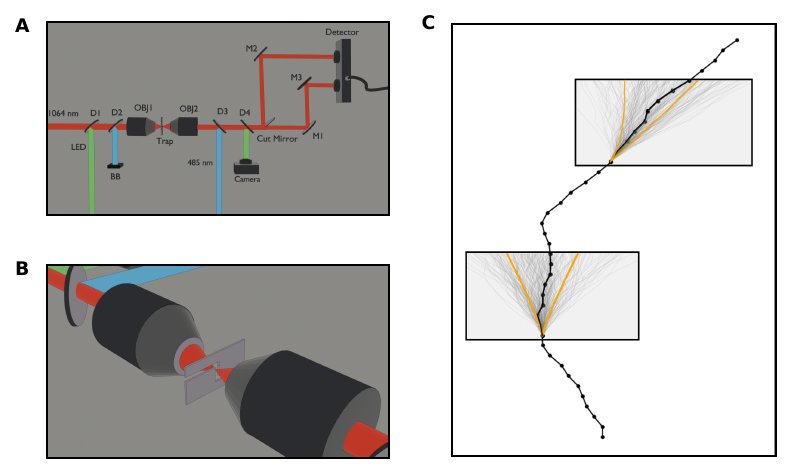}
\caption{\label{graphics}\textbf{Short time Brownian motion detection scheme.}\label{experimetal_setup} (\textbf{A}) Diagram of optical trapping and detection setup. Counter-propagating beams are focused through a microfluidic chamber forming an optical tweezer and trapping the particle. The out-coming infrared beam is re-collimated and split spatially with a D-shaped cut mirror. Each half of the beam is sent to a port on a balanced photodetector which monitors the particle's position. (\textbf{B}) Close up of the microfluidic chamber with z-shaped channel and focused beam passing through. (\textbf{C}) Time trace data with time on the vertical and position on the horizontal axis. Two points are chosen that fall within the conditioning tolerance around 0 velocity and +1 standard deviation of the velocity, respectively. Overlaid on these initial points are sub-traces of other parts of the time trace that fall within the same tolerance along orange lines indicating 1 standard deviation from the mean trajectory.}
\end{figure}

\begin{figure} 
	\centering
	\includegraphics[width=0.7\textwidth]{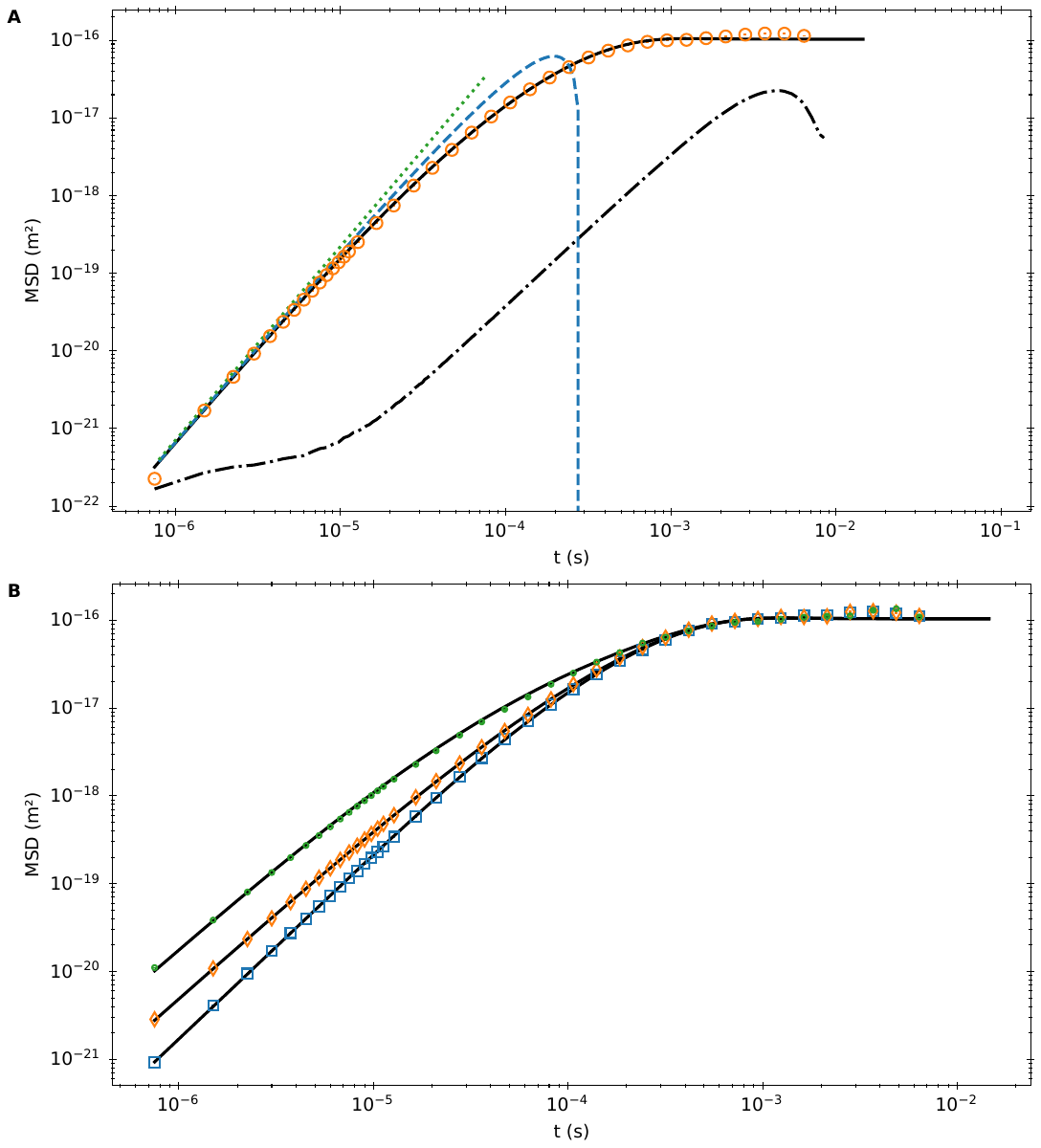} 

	\caption{\textbf{MSDs for a Brownian particle for specific initial velocities.}\label{set_velos_graph} (\textbf{A}) MSD built from trajectories with the particle beginning close to rest. The orange circles are experimental data points. The solid black line is the hydrodynamic theory. The blue dashed line corresponds to equation \ref{two_term_expansion}. The green dotted line is the $t^{5/2}$ scaling term, which the experimental data collapses onto at short times. The dot-dashed line corresponds to the MSD found with the same velocity conditioning and no particle in the trap. (\textbf{B}) MSD curves for three different initial velocities. The blue squares, orange diamonds, and green circles refer to initial velocities of $0.5$, $1$, and $2$ times the velocity standard deviation respectively. The black lines are the theory. Error bars for both graphs are calculated using the method outlined in the Supplementary Materials.} 
\end{figure}

Using a custom-built high-powered balanced photodetector and split beam detection pioneered in \cite{kheifets_observation_2014}, we track the position of a barium titanate microsphere (diameter $6.8 \pm 0.2$ $\mu$m) optically trapped in acetone well below its momentum relaxation time. A schematic of the experimental setup can be seen in Fig. \ref{experimetal_setup}. 
With an eighth order finite difference, we estimate the velocity of the particle close to its true instantaneous velocity. We find moments in time when the velocity is less than 1 percent of the experimentally measured velocity standard deviation. We then average the ensemble of sub-traces beginning at these moments to build an MSD with an initial velocity close to 0. The result can be seen in Fig. \ref{set_velos_graph}. We find that the experimental curve converges with equation \ref{two_term_expansion} at short times, demonstrating the short-time MSD scaling enforced by the thermal force correlation function. Note that since the velocity is calculated via a finite differencing scheme, the first data point is set by the magnitude of the conditioned velocity and the laser noise of the system. The full theoretical curve is derived by analyzing the problem in the Laplace domain. \par 

Solving the hydrodynamic GLE in the Laplace domain \cite{clercx_brownian_1992, lutz_fractional_2001, bakalis_hydrodynamic_2023, burov_fractional_2008} as opposed to the Fourier domain enables the inclusion of initial conditions for $v(0)$ and $x(0)$ and affords limited tools to deal with the history's effect on future dynamics \cite{di_terlizzi_explicit_2020}. In the following analysis, we allow for an arbitrary initial velocity and set $x(0)=0$ for simplicity. We separate the Basset-Boussinesq force at $t=0$ and integrate the history term by parts to obtain

\begin{multline}  
    M \ddot{x}(t) = -\gamma \dot{x}(t) - \gamma v(0)\sqrt{\frac{\tau_f}{\pi t}}+ \gamma\sqrt{\frac{\tau_f}{\pi}}\int_{-\infty}^0\frac{\dot{x}(\tau)}{2(t - \tau)^{3/2}}d\tau \\ -\gamma\sqrt{\frac{\tau_f}{\pi}}\int_{0}^t\frac{\ddot{x}(\tau)}{\sqrt{t - \tau}}d\tau -Kx(t) + R(t) \, .
\end{multline}
The history integral's boundary term cancels with an initial condition term in the Laplace domain, and the remaining integral is left as the history effect, which we refer to as the history force. We introduce $B(t)$ as the Green's function of the system and find the solution for $x(t)$ as a function of $v(0)$, the thermal force, and history force.

\begin{equation}
\label{x_solution_no_x0}
x(t) = Mv(0)B(t)+\gamma\sqrt{\frac{\tau_f}{\pi}}\int_{0}^{t}B(t-t') \int_{-\infty}^0\frac{\dot{x}(\tau)}{2(t' - \tau)^{3/2}}d\tau dt' +\int_{0}^{t}B(t-t')R(t')dt' \, .
\end{equation}

When squaring equation \ref{x_solution_no_x0} to find the MSD, we get terms that involve correlations between the thermal force and history force conditioned on a specific initial velocity. Assuming joint-Gaussianity implies that these conditional correlations can be calculated with knowledge of the system's full covariance matrix \cite{joram_soch_statproofbookstatproofbookgithubio_2025}. While the conditioned history force can be estimated via the particle's equilibrium VACF, we are left to find an expression for $\langle R(t)v(0)\rangle$. We start from the hydrodynamic GLE of a free particle:

\begin{equation}
    \label{parted_GLE}
    M\dot{v}(t) = -\gamma v(t) - \gamma v(0) \sqrt{\frac{\tau_f}{\pi t}} +\gamma \sqrt{\frac{\tau_f}{\pi}} \int_{-\infty}^0\frac{v(t')}{2({t -t'})^{3/2}}dt' -\gamma \sqrt{\frac{\tau_f}{\pi}} \int_0^t\frac{\dot{v}(t')}{\sqrt{t -t'}}dt'+ R(t) \, .
\end{equation}

Following the analysis presented in \cite{giona_dynamic_2025, tothova_statistical_2016}, we multiply each side of equation \ref{parted_GLE} by $v(0)$, take an average over equilibrium conditions, and perform a Laplace transform. This yields the result

\begin{equation}
    \label{equal_laplace}
    \tilde{C}_{vv}(s) = \tilde{\mu}(s)\big[k_BT +\mathscr{L}[\langle R(t)v(0) \rangle + \gamma\sqrt{\frac{\tau_f}{\pi}}\int_{-\infty}^0\frac{\langle v(0)v(t') \rangle }{2({t -t'})^{3/2}}dt']\big] \, ,
\end{equation}
where $\mathscr{L}$ denotes the Laplace transform, $C_{vv}$ is the VACF, and $\tilde{\mu}(s)$ is the admittance of the system. As a consequence of the fluctuation-dissipation theorem \cite{kubo_fluctuation-dissipation_1966}, it follows that 
\begin{equation}
    \label{fluc_dis_result}
    \tilde{C}_{vv}(s) = k_BT\tilde{\mu}(s) \, .
\end{equation}
Accordingly, from equations \ref{equal_laplace} and \ref{fluc_dis_result} we find
\begin{equation}
    \label{equil_RV_condition}
    \langle R(t)v(0) \rangle = - \gamma\sqrt{\frac{\tau_f}{\pi}}\int_{-\infty}^0\frac{\langle v(0)v(t') \rangle }{2({t -t'})^{3/2}}dt' \, .
\end{equation}
Equation \ref{equil_RV_condition} is a consequence of the fluctuation-dissipation theorem and holds in equilibrium. It defines the covariance between the particle's history and future thermal force. An analogous expression has been shown in \cite{london_forcevelocity_1977} for the GLE with velocity instead of acceleration appearing in the memory kernel and matches the early formulations of GLEs by Kubo. Additionally, in a harmonic trap, correlations between the thermal force and position yield
\begin{equation}
    \label{equil_RX_condition}
    \langle R(t)x(0) \rangle = - \gamma\sqrt{\frac{\tau_f}{\pi}}\int_{-\infty}^0\frac{\langle v(t')x(0) \rangle }{2({t -t'})^{3/2}}dt' \, .
\end{equation} 
Thus, the covariance matrix and conditioned expectation values for the velocity, position, history force, and thermal force can be calculated.

We find that the conditioned MSD can be expressed in terms of only the equilibrium thermal force correlation and the initial conditions, $x(0)$ and $v(0)$ (see Supplementary Materials). The MSD for a particle beginning at the center of the trap is 


\begin{equation}
    \label{abstract_msd}
    \operatorname{MSD}[t]= M^2v(0)^2B(t)^2 +
      \int_0^{t} B(t-\tau)\int_0^{t}B(t-t')\langle R(\tau)R(t')\rangle dt' d\tau \, .
\end{equation}
Analogous to \cite{clercx_brownian_1992, fox_gaussian_1978}, the thermal force integral can be solved in the Laplace domain, yielding an analytic expression for the conditioned MSD.

To compare with experimental data, we identify the velocity of our particle by a ratio with the experimentally accessible velocity standard deviation and then compare to the corresponding ratio from the theoretical Maxwell-Boltzmann distribution. Our experimental velocity standard deviation is 98\% of the theoretical one, with an SNR of 12 dB. Then, we select any moment in time when our particle has a velocity within one percent of the experimental velocity standard deviation from the desired initial condition. To obtain better statistics, we leave the position degree of freedom thermally distributed. Traces beginning at these times then form the ensemble from which we calculate the experimental MSD curve. A comparison of these experimental MSDs with their theoretic counterparts can be seen in Fig. \ref{set_velos_graph}. The strong agreement between the two indicates a correct analysis of the interplay between the nonequilibrium thermal force and history terms.\par 

Analysis of the mean trajectory provides an even stronger test for the cross-correlation. When the initial velocity is predetermined, the terms in equation \ref{x_solution_no_x0} involving the initial velocity become deterministic, yielding a nonzero mean trajectory of our Brownian sphere. We can write this mean trajectory as 

\begin{equation}
    \label{mean_trajectory_eq}
    \langle x(t)\rangle  = Mv(0) B(t) + \int_0^tB(t-t')\big(\langle R(t')|v(0)\rangle + \int_{-\infty}^0\frac{ \langle v(\tau)|v(0)\rangle }{2(t'-\tau)^{3/2}}d\tau \big) dt' \, .
\end{equation}
Due to equation \ref{equil_RV_condition}, however, the two terms in the integral cancel precisely (see Supplementary Materials) yielding a simple form for the mean trajectory of
 
\begin{equation}
    \label{mean_trajectory_canceled}
    \langle x(t)\rangle  = Mv(0) B(t) \, .
\end{equation}
\begin{figure}
\centering
\includegraphics[width=0.725\textwidth]{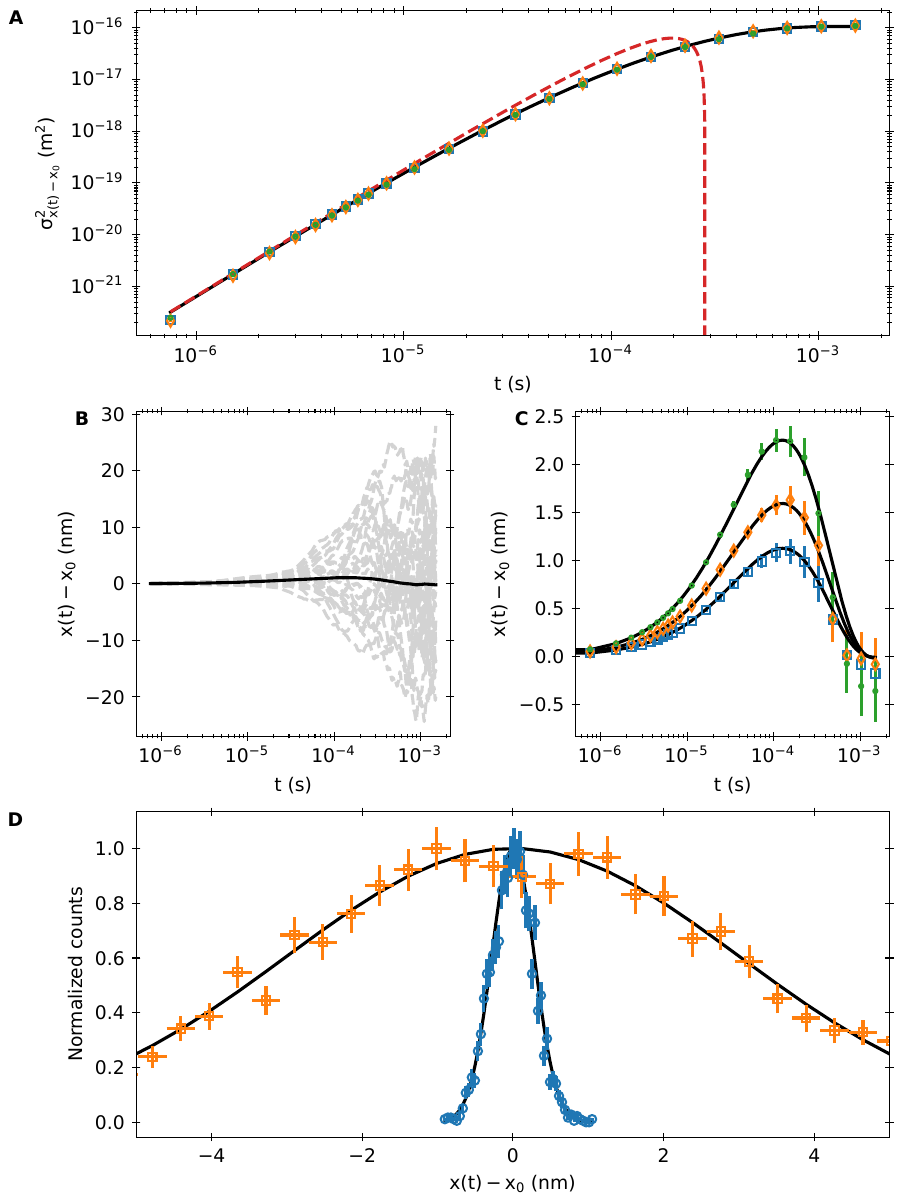}
\caption{\label{fluctuating_graph} \textbf{Analysis of mean and fluctuations around the mean trajectories for the Brownian particle.} (\textbf{A}) Fluctuations around the mean trajectory for 3 different initial velocities. The blue squares, orange diamonds, and green circles correspond to initial velocities with squared values $0.5 \langle v^2 \rangle$, $\langle v^2 \rangle$, $2\langle v^2 \rangle$ respectively. The black line is the full theory and the red dashed line is the short time expansion from equation \ref{two_term_expansion}. (\textbf{B}) Trajectories for the Brownian particle with initial velocity squared of $\langle v^2 \rangle$. The black line is the mean trajectory, and the gray dashed lines are individual trajectories. (\textbf{C})  Mean trajectory for different initial velocities. The blue squares, orange diamonds, and green circles correspond to initial velocities with squared values $0.5 \langle v^2 \rangle$, $\langle v^2 \rangle$, $2\langle v^2 \rangle$ respectively. Solid lines are the theory from equation \ref{mean_trajectory_eq}. Error bars for panels \textbf{A} and \textbf{C} are calculated using the method outlined in the Supplementary Materials. (\textbf{D}) Distribution of fluctuations around the mean trajectory for an initial velocity squared of $\langle v^2 \rangle$. The orange squares are for a time of 75 $\mu$s and the blue circles for a time of 7.5 $\mu$s. Black lines are Gaussian curves with standard deviation set by the zero velocity MSD.}
\end{figure}
We can access this mean trajectory experimentally by conditioning in an identical manner. Experimental results with curves for equation \ref{mean_trajectory_canceled} can be seen in Fig. \ref{fluctuating_graph}. While limited statistics yield larger uncertainties at longer times, the experimental data appears to agree well with the theoretical curves. Therefore, we see that when our initial velocity is well-defined, the thermal force acquires a nonzero mean value that decays back to zero. Furthermore, this nonzero mean directly cancels the remaining history term as a consequence of the fluctuation-dissipation theorem.\par
It is clear that the fluctuations grow fast around the mean. To describe these fluctuations, note they are governed by the terms remaining in equation \ref{abstract_msd} when the deterministic terms are removed. Therefore, when selecting for specific initial velocity and position, we expect 

\begin{equation}
    \langle (x(t) - \langle x(t) \rangle )^2\rangle = \int_0^{t} B(t-\tau)\int_0^{t}B(t-t')\langle R(\tau)R(t')\rangle dt' d\tau \, .
\end{equation}
Thus, the fluctuations should only depend on the equilibrium correlation function of $R(t)$. 

To get better statistics, we condition only over the initial velocity of the particle, so we also expect fluctuations due to the initial position. However, the thermal forcing should still dominate at short times. In fact, since we expect the fluctuations to only depend on the equilibrium form of $R(t)$, the temporal growth of these fluctuations around the mean trajectory should look identical to the zero velocity super-ballistic MSD from Fig. \ref{set_velos_graph} \cite{tateishi_different_2012}. Indeed, as seen in Fig. \ref{fluctuating_graph}, by averaging the fluctuations around the mean trajectory for three different initial speeds, all three collapse onto the zero velocity curve. Then, to within experimental uncertainties, we see that the fluctuations around the mean value are equivalent to forcings by a stochastic force with statistics identical to the equilibrium thermal force. Note, the true thermal force does not take its equilibrium value. Rather, the nonequilibrium component of the thermal force is exactly canceled by the nonequilibrium state of the Basset force. A summary of this analysis is provided by an effective GLE for equilibrium particle dynamics with a set initial velocity. For a given initial velocity $v(0)$, the future dynamics of a free particle are modeled by

\begin{equation}
        \label{eff_GLE}
        M\dot{v}(t) = -\gamma v(t) -\gamma \sqrt{\frac{\tau_f}{\pi}} \int_0^t\frac{\dot{v}(t')}{\sqrt{t -t'}}dt' -  \gamma \sqrt{\frac{\tau_f}{\pi}}\frac{v(0)}{\sqrt{t}} +R_{eff}(t) - K x(t)\, ,
\end{equation}
where $R_{eff}(t)$ is an effective thermal force that has equilibrium correlation properties defined in equation \ref{equilibrium_thermal_corr}. Note this form matches the equation presented in \cite{lutz_fractional_2001, procopio_modal_2023}. Here, we have shown how this initial value representation can be directly derived from the equilibrium, infinite past representation and demonstrated its experimental validity through both the mean trajectory and fluctuational analysis. Furthermore, the Langevin condition $\langle R_{eff}(t) v(0) \rangle = 0$ is recovered similarly to the analysis for GLEs shown in \cite{giona_dynamic_2025, procopio_modal_2023} and in line with the original formulation of the fluctuation-dissipation theorem and initial value representation of GLEs \cite{kubo_fluctuation-dissipation_1966}. As in Kubo's original work, the Langevin condition only holds for this effective thermal force; the true thermal force is dependent on $v(0)$. As observed in \cite{london_forcevelocity_1977, tothova_colour_2013}, there is no violation of causality; the past thermal force influences $v(0)$ dynamically and $R(t)$ is correlated with its own past, so $v(0)$ and $R(t)$ are correlated as well. Note the boundary term is still necessary to accurately model the future dynamics.
\par
In Fig. \ref{fluctuating_graph}, we also plot the distribution of the fluctuations for a singular initial velocity. Since these fluctuations are driven primarily by the thermal force at short times, they offer an opportunity to probe thermal force statistics beyond the second moment. If the thermal force is Gaussian, the mean trajectory and the second moment set by the fluctuations would completely determine our trajectories' statistics. However, recent theoretical works argue that the thermal force may not be Gaussian \cite{giona_extended_2022, pezzotti_fluid-particle_2024}. For this single velocity, there are not enough statistics to see much of the tails. \par

\begin{figure}
\centering
\includegraphics[scale=0.8]{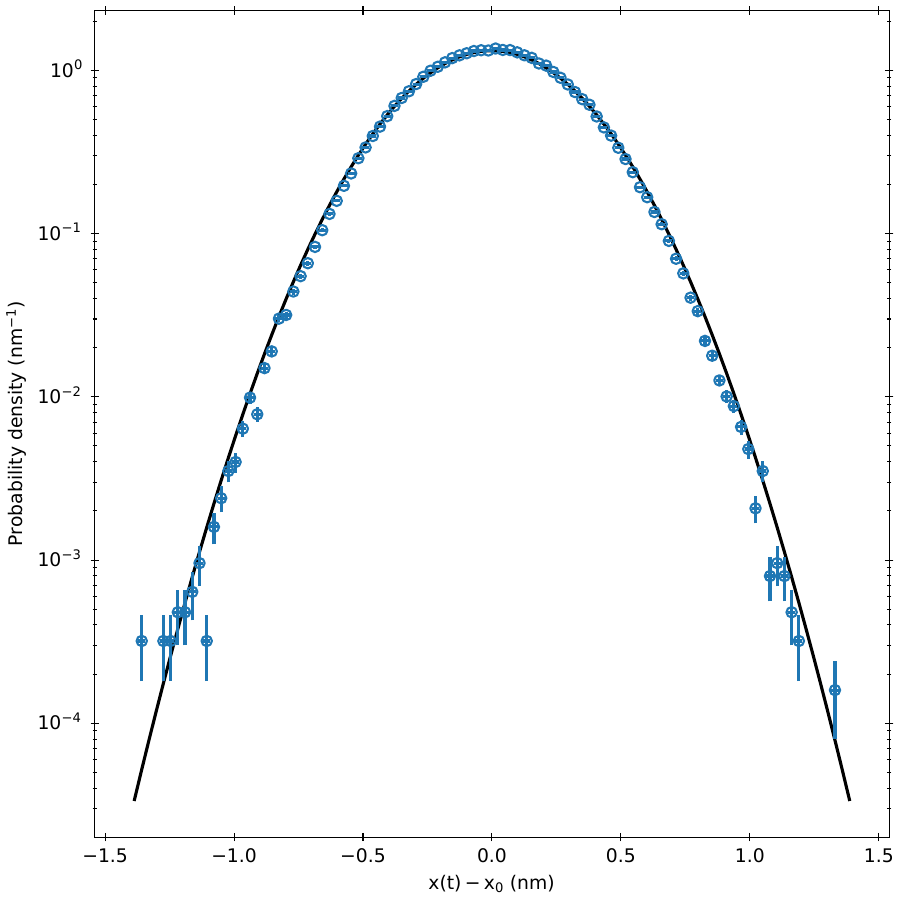}

\caption{\label{histogram} \textbf{Histogram for fluctuations around the mean trajectory.} Built for initial speeds between minus one and one velocity standard deviation. The variations are taken 7.5 $\mu$s after the starting trajectory. The variation at these short times is caused by the thermal force, and therefore is a good test for its finer stochastic properties. We find a standard deviation 97\% the value predicted by the theory, which could result from the density uncertainty of the sphere or finite differencing effects on the fractal-like time series.}
\end{figure}

To improve the statistics, we divide all velocities lying within 1 standard deviation of the mean into bins with widths 2 percent of the experimental velocity standard deviation. A limit of 1 standard deviation is chosen, as to ensure enough statistics to accurately estimate the mean trajectory. We then find the fluctuations around the mean trajectory for each bin. The histogram of all fluctuations $7.5$ $\mu$s after the initial point is shown in Fig. \ref{histogram}. The observed fluctuations around the mean follow a Gaussian distribution well, although a much finer temporal resolution may still find deviations. We find a standard deviation 97\% the value predicted by our super-ballistic theory. Therefore, the Gaussian assumption for the thermal force distribution is valid for the timescales of interest.\par
\begin{figure}
\centering
\includegraphics[scale=1]{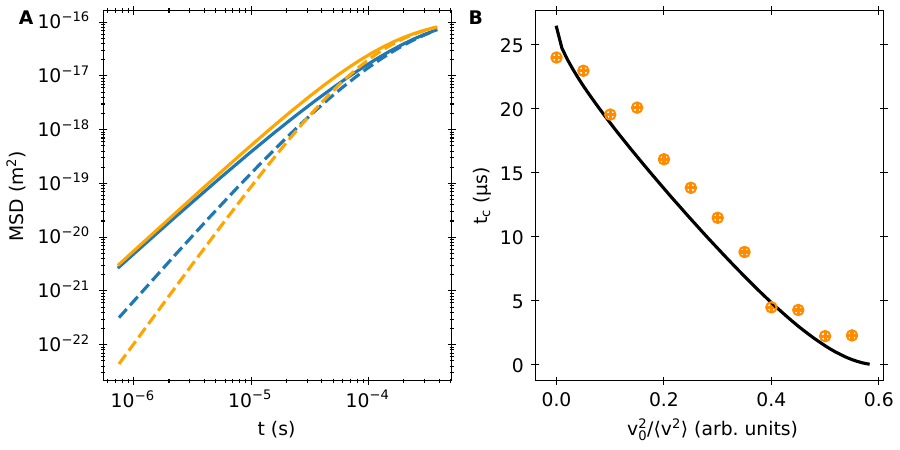}

\caption{\label{crossings} \textbf{Comparison of white noise and hydrodynamic Langevin equations.} (\textbf{A}) MSDs for hydrodynamic and white noise Langevin equations with different initial conditions. The hydrodynamic curves are blue, while the white noise curves are orange. The solid lines are MSDs taken with thermal initial conditions, while the dashed lines assume an initial velocity of 0. (\textbf{B}) Crossing times as a function of initial velocity. The solid black line is the crossing of the two analytic curves. The orange circles are the experimental times when a crossing occurs. The vertical error bars are set by the time step between data points. The horizontal error bars are set by the width of the velocity bins used in the analysis.}
\end{figure}

\FloatBarrier
The presence of the hydrodynamic colored noise raises the short time MSD from $t^3$ for a Brownian particle driven by white noise to $t^{5/2}$. Therefore, we see increased transport in the hydrodynamic case at short times in comparison to a purely white-noise description. At longer times, the presence of the Basset-Boussinesq force slows down transport, so the hydrodynamic and white noise MSDs cross at some crossing time $t_c$ as seen in Fig. \ref{crossings}. However, when we take thermal initial conditions for our velocity, the short time scaling of the MSD is dominated by the initial velocity of the particle, and we see no occurrence of a crossing-time. To explore this behavior, we find $t_c$ as a function of the initial velocity. As seen in Fig. \ref{crossings}, the crossing time is a decreasing function of initial velocity. This decreasing behavior continues until some critical velocity, where now the initial velocity obscures acceleration from the colored thermal force. The crossing time analysis separates two qualitatively different short-time regimes. For small initial velocities, increased transport from the colored noise can temporarily dominate transport. For larger initial velocities, decreased transport from the Basset term is the dominant hydrodynamic effect. \par

\section{Discussion}
A natural consequence of this work regards the definition of equilibrium in systems with large fluctuations. All the data presented in this study were taken in equilibrium, in the sense that the time averaged quantities for the system match equilibrium predictions. Nevertheless, by modifying how the averaging is done, we find ensembles of trajectories where the microsphere does not possess its equilibrium properties. This allows direct observation of relaxation back to equilibrium with no changes to the physical apparatus. We can use data collected in equilibrium situations to study nonequilibrium processes. The nonequilibrium states we have access to are limited both by practical and fundamental limits. Practically, we can only probe so far into the distributional tails with a realistic amount of data collection time. Fundamentally, in these non-Markovian systems, we see the state of the fluid and the state of the particle are strongly correlated, so in equilibrium defining the particle's initial statistics also defines the fluid's, and we cannot easily decouple them. \par
Nevertheless, this conditioning method demonstrates the importance of the non-Markovian interactions between the particle and the fluid. Conservation laws in incompressible fluids, governed by Navier-Stokes, demand increased fluctuations and the $t^{5/2}$ scaling at short times. Of course, the existence of the crossing time $t_c$ for the zero initial velocity trajectories also indicates that this enhanced stochasticity is transient; eventually the memory of the fluid conserves more information about the particle's past than the additional thermal force destroys. Finally, the analysis of the history term and experimental confirmation of the effective GLE demonstrates the coupling of the fluid-particle system and verifies a theoretical tool for future analysis.

\section{Materials and Methods}

Fig. \ref{graphics} shows a simplified schematic of our experiment. We trap a barium titanate microsphere ($\rho = 4200 \pm 200$ kg/$m^3$) in acetone ($\eta = 0.32$ mPas, $\rho_f = 790$ kg/$m^3$) at a temperature of 293 K with two counter-propagating laser beams forming an optical tweezer. The two beams (1064 nm, ~300 mW and 485 nm, ~200 mW) are focused by two oil immersion microscope objectives through a home-built microfluidic flow cell. The flow cell is constructed by cutting a channel into parafilm, and placing this channel between two glass cover-slips. The cover-slips are then placed in a laminating sheet with optical access windows cut out and capillary tubes attached. This system is sealed by lamination and fluid is introduced through the capillary tubes.\par  The thermal motion of the microsphere is resolved by spatially halving the re-collimated infrared laser light, which carries information about the particle position in its antisymmetric component \cite{madsen_ultrafast_2021}. The difference in power in the two halves of the beam varies linearly with the microsphere position in the vicinity of the trap's center and is measured using a custom high-power split beam detector. The custom detector is modeled off the original high-powered detector used to resolve the Maxwell Boltzmann distribution of a particle \cite{mo_testing_2015}. After losses through the trap, we have approximately 50 mW of laser power to apply to each photodiode. The output voltage of the detector is measured on a 16-bit digitization card, taking $2^{24}$ samples at 200 MSamples/s. The bandwidth of the measurement is limited by the laser’s noise which sets the time resolution of the trace to $750$ ns. \par
The detector has a built-in high-pass filter on its trans-impedance stage to prevent saturation of the electronics by low-frequency noise and Brownian motion. In past experiments, the low frequency noise has been overcome by manually whitening the PSD of the particle \cite{mo_highly_2019} or high pass filtering the data and only looking at the high-frequency ballistic motion \cite{madsen_ultrafast_2021}. However, since we are interested in time traces from short times up through equilibration, we measure the response of this filter and then invert its effects on our experimental data with a Tikhonov regularization scheme to limit the effects from the low frequency electronic noise floor. \par
\begin{figure} 
	\centering
	\includegraphics[width=0.9\textwidth]{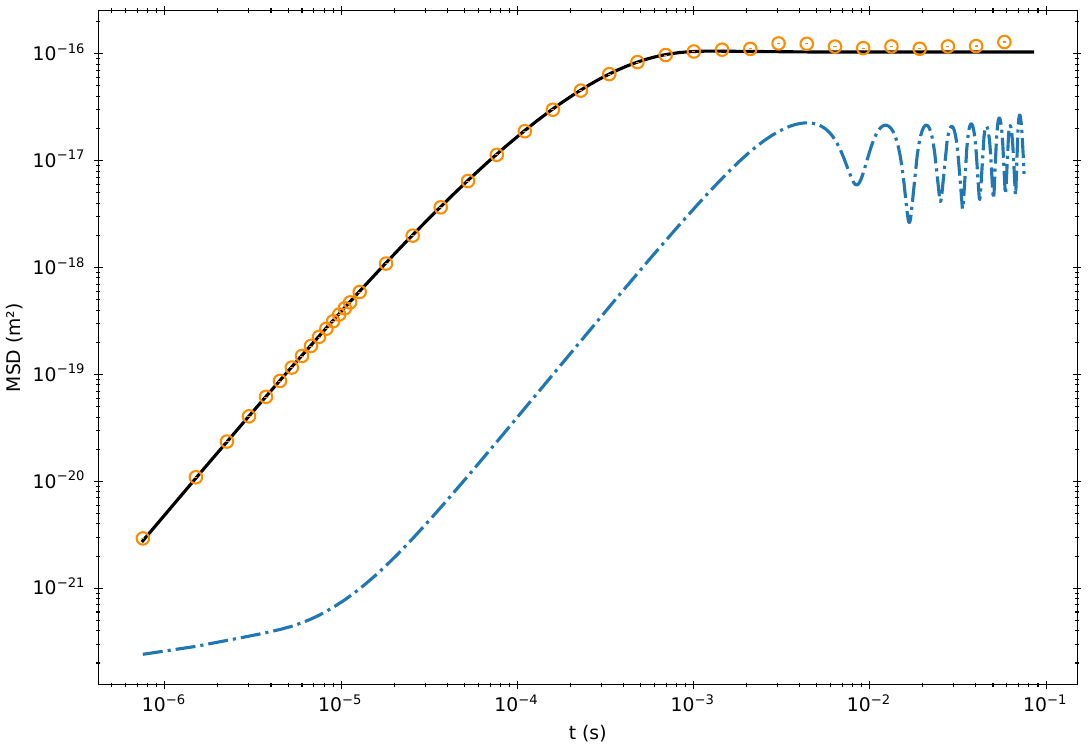} 

    \caption{\label{equilibrium_plot} \textbf{Experimental MSD for trapped Brownian particle.} MSD built by averaging displacements starting from every point in the time trace. The orange dots are the experimental data points. The black line is a fit to the equilibrium MSD from \cite{clercx_brownian_1992}, and the blue dot-dashed line is the MSD from the system with no sphere trapped. The fit is performed up to 1.2 ms to limit the impact of low frequency noise and regularization scheme. The error bars are calculated using the blocking method outlined in \cite{flyvbjerg_error_1989}.}
\end{figure}

We calibrate the experimental voltage time trace by performing a log-spaced least-squares fit to the equilibrium hydrodynamic MSD from six independent traces, each of length 111,848, as shown in Fig. \ref{equilibrium_plot}. We fit the experimental MSD for the particle diameter, the trap strength, and the voltage to position detector calibration factor. The mass is assumed to be equal to $4/3 \pi a^3 (\rho +\rho_f/2)$ to account for the added mass. This analysis yields values for the particle diameter, trap strength, and detector calibration coefficient of $6.8 \pm 0.2$  $ \mu$m, $78 \pm 4$  $\mu $N/m, and $29 \pm 1 $ mV/nm respectively. To further reduce the effects of low frequency noise and the regularization procedure, we fit the MSD up to a maximum time of $1.2$ ms. The reported uncertainties correspond to the uncertainty between the 6 fits. We apply an eighth-order finite differencing scheme to our position traces, resulting in a velocity variance of 98\% the value of the associated Maxwell-Boltzmann distribution with an SNR of 12 dB. We therefore resolve much of the velocity correlation of the Brownian particle. \par
To determine the theoretical crossing times, we use our analytic solution for the conditional hydrodynamic MSD and compare it to an analytic solution of the white-noise Langevin equation (see Supplementary Materials). The crossing time is the intersection of these two curves. For the experimental results, we generate an interpolated function from the experimental MSD curve using the scipy.interpolate.interp1d() function. We then numerically determine when this interpolated function crosses paths with the white noise solution. While throughout we have been focusing on the effect of Basset force and colored thermal noise, the hydrodynamics also re-normalizes the mass of the particle leading to the so-called added mass effect. Therefore, we subtract the added mass off the experimentally determined mass when calculating the white-noise curves. Note while this modifies the mass of the particle, we still initialize both the white-noise model and the hydrodynamic model with same velocity with $v^2/\langle v_0^2 \rangle $ set by the effective mass (the bare mass of the particle plus the added mass from the surrounding fluid). Note for nonzero velocity, the leading order term will always go like $t^2$, and at short enough times the motion will behave ballistically. At these experimentally inaccessible times, we can have another crossing time where the hydrodynamic curve crosses the white noise curve. We restrict our attention to the experimentally accessible final crossing time.



\clearpage 

%
\bibliography{references} 
\bibliographystyle{sciencemag}

%
%
%
%
%
%

\newpage
\paragraph*{Acknowledgments:}
We thank Massimiliano Giona, Giuseppe Procopio, and Chiara Pezzotti for helpful conversations on the hydrodynamic theory.
\paragraph*{Funding:}
JB is supported by the NSF Graduate Research Fellowship Program under Grant No. DGE 2137420. MCT is supported by the National Defense Science and Engineering Graduate (NDSEG) Fellowship. This research is supported by the Sid W. Richardson Foundation.
\paragraph*{Author contributions:}
Conceptualization: MGR, MCT, JB, Methodology: JB, MCT, Software: MCT, JB, Validation: JB, MCT, Formal Analysis: JB, MCT, Investigation: MCT, JB, Resources: MGR, Data curation: MCT, JB, Writing – original draft preparation: JB, MCT, Writing – review and editing: MGR, JB, MCT, Visualization: MCT, JB, Supervision: MGR, Project Administration: MGR, Funding acquisition: MGR

\paragraph*{Competing interests:}
The authors declare no competing interests.

\paragraph*{Data and materials availability:}
The data for this study have been deposited in the database Dryad at https://doi.org/10.5061/dryad.pvmcvdnz4.

\newpage


\renewcommand{\thefigure}{S\arabic{figure}}
\renewcommand{\thetable}{S\arabic{table}}
\renewcommand{\theequation}{S\arabic{equation}}
\renewcommand{\thepage}{S\arabic{page}}
\setcounter{figure}{0}
\setcounter{table}{0}
\setcounter{equation}{0}
\setcounter{page}{1} 


\begin{center}
\section*{Supplementary Materials for\\ \scititle}

Jason Boynewicz$^{\ast\dagger}$,
Michael C. Thumann$^\dagger$,
Mark G. Raizen\\ 
\small$^\ast$Corresponding author. Email: jrb8338@my.utexas.edu\\
\small$^\dagger$These authors contributed equally to this work.
\end{center}

\subsubsection*{This PDF file includes:}
Supplementary Text\\
Figure S1

\newpage


\section{Supplementary Text}
\subsubsection{High-Powered Detector and transfer function}
\begin{figure}
\centering
\includegraphics{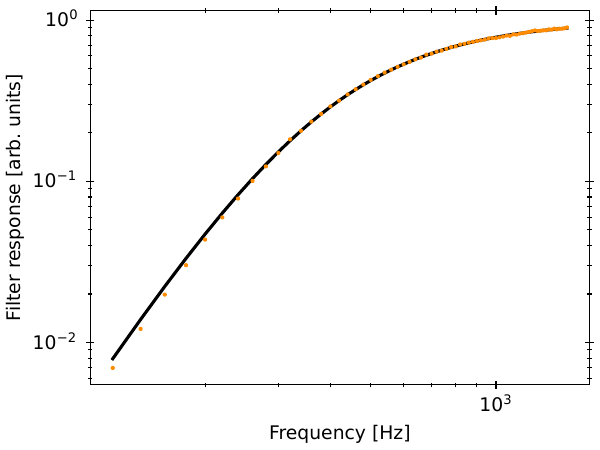}%

\caption{\label{transfer}\textbf{Balanced detector characterization.} Plot of the amplitude response for the high pass filter on the high-powered balanced photodetector. The data are shown with the orange circles. The two-pole high pass filter fit is included with the solid black line.}
\end{figure}

We use a custom-built high-powered detector capable of withstanding 100 mW of laser power on each port of the detector. A second-order high-pass trans-impedance filter converts the photocurrent difference from the two detectors into a voltage signal. The high-pass filter built into the detector is necessary in order to apply the large incident powers onto the detector. Without the filter, low-frequency motion of the sphere and noise saturate the operational amplifier well before all incident laser power is used. To account for the filter, we empirically determine the transfer function of the detector by modulating the laser at known frequencies and monitoring the response of the detector. Since the laser itself has its own frequency-dependent response, the modulation was also monitored on a photodetector with a flat frequency response (Thorlabs PDB425C). We then fit a two-pole high-pass filter model to the empirical results. The resulting high-pass filter frequency response can be seen in Fig. \ref{transfer}. \par
Since the high-pass filter becomes singular at zero frequency, we face an inversion problem due to the low-frequency electronic noise sources in the system. To prevent large amounts of anomalous low-frequency noise while still inverting the high-pass filter, we apply a Tikhonov regularization scheme on the data. To do so, we Fourier transform our time trace, and then multiply the Fourier components by 
\begin{equation}
    \frac{H^*(f)}{|H(f)|^2 + |H(40)|^2} \, ,
\end{equation}
where $H(f)$ is the transfer function of the high-pass filter in the detector. When the frequency approaches 40 Hz, the presence of the second term in the denominator prevents a singularity.\par
To ensure the regularization scheme has minimal impact on the extraction of the trap's physical parameters, we perform our same fitting procedure for several different cut frequencies between 20 Hz and 70 Hz. We find minimal dependence on the cut frequency with negligible changes in the radius and volts to meter conversion and changes in the strength of the trap on the order of 3.2\%.



\newpage

\subsubsection{Expansion of 0 Velocity MSD}
For a free particle beginning at rest with no previous history of the fluid, Clercx and Schram \cite{clercx_brownian_1992} derived the VACF to be of the form
\begin{equation}
    \langle v(t_1)v(t_2)\rangle = k_BT[\Lambda(|t_1 - t_2|) - M\Lambda(t_1)\Lambda(t_2)] \, ,
\end{equation}
where 
\begin{align}
    \Lambda(t) &= \frac{1}{M(b-a)}[b e^{b^2t}\operatorname{erfc}[b\sqrt{t}] - ae^{a^2 t}\operatorname{erfc}[a \sqrt{t}]] \\
    z&= 6 \pi a^2 \sqrt{\rho_f \eta}\\
    a &= \frac{z}{2M}(1+\sqrt{1 -\frac{4 \gamma M}{z}}) \\
    b &= \frac{z}{2M}(1 - \sqrt{1 -\frac{4 \gamma M}{z}}) \, .
\end{align}
We expand this expression in powers of $\sqrt{t}$ to second order. Doing so yields an approximate short time expression for the VACF of
\begin{equation}
    \langle v(t_1)v(t_2)\rangle \approx \frac{k_BT}{M^2}( \frac{2 z}{\sqrt{\pi}}(\sqrt{t_1} + \sqrt{t_2} - \sqrt{|t_1-t_2|}) + \frac{2}{M}(M \gamma - z^2)\min{[t_1, t_2]} - \frac{4 z^2}{{M\pi}} \sqrt{t_1t_2}) \, .
\end{equation}
The MSD can be found by integrating the VACF with respect to $t_1$ and $t_2$ yielding the result quoted in the main paper for the short time super-ballistic motion of
\begin{equation}
    \operatorname{MSD}[t] = \frac{2 k_B T}{3M \tau_p} [\frac{12}{5 }\sqrt{\frac{\tau_f}{\pi}}t^{5/2} + \beta t^3] \, .
\end{equation}

\subsubsection{Solving the Hydrodynamic GLE}
\par First, we separate the Basset-Boussinesq force at bound $t=0$ and integrate the historical Basset-Boussinesq term by parts. The GLE becomes 
\begin{equation}
    M \ddot{x}(t) = -\gamma \dot{x}(t) - \gamma \sqrt{\frac{\tau_f}{\pi}}\frac{v(0)}{\sqrt{t}} + I(t) -\gamma\sqrt{\frac{\tau_f}{\pi}}\int_{0}^t\frac{\ddot{x}(\tau)}{\sqrt{t - \tau}}d\tau -Kx(t) + R(t) \, ,
\end{equation} 
where we denote the history force
\begin{equation}
    I(t) = \gamma\sqrt{\frac{\tau_f}{\pi}}\int_{-\infty}^0\frac{\dot{x}(\tau)}{2(t - \tau)^{3/2}}d\tau \, .
\end{equation}
This procedure allows the boundary term to cancel the initial velocity term of the forward Basset-Boussinesq convolution in Laplace space. We then solve the equation by transforming to Laplace space to obtain
\begin{multline}
    M[s^2\tilde x(s)-sx(0)-v(0)] = -\gamma[s\tilde{x}(s)-x(0)] - \gamma \sqrt{\frac{\tau_f}{s}}v(0) +\tilde{I}(s) \\ -\gamma\sqrt{\frac{\tau_f}{s}} [s^2\tilde{x}(s)-sx(0)-v(0)] -K\tilde{x}(s) + \tilde{R}(s) \, .
\end{multline}
We solve for $\tilde{x}(s)$ in Laplace space and invert to get a function for $x$ in the time domain and find
\begin{multline}
\label{x_solution}
x(t) = Mx(0)A(t)+Mv(0)B(t)+\gamma x(0)B(t) + \gamma \sqrt{\tau_f}x(0)D(t)+ \\ \int_{0}^{t}B(t-t')I(t')dt' +\int_{0}^{t}B(t-t')R(t')dt' \, ,
\end{multline}
where $A(t)$, $B(t)$, and $D(t)$ are defined in \cite{clercx_brownian_1992} as
\begin{equation}
A(t) = \frac{1}{M} \sum_{i=1}^{4} 
\frac{q_i^3 \, e^{q_i^2 t} \, \operatorname{erfc}(q_i \sqrt{t})}
{\displaystyle\prod_{j=1, j \ne i}^{4} (q_i - q_j)} 
\end{equation}
\begin{equation}
B(t) = \frac{1}{M} \sum_{i=1}^{4} 
\frac{q_i \, e^{q_i^2 t} \, \operatorname{erfc}(q_i \sqrt{t})}
{\displaystyle\prod_{j=1, j \ne i}^{4} (q_i - q_j)} 
\end{equation}

\begin{equation}
D(t) = \frac{1}{M} \sum_{i=1}^{4} 
\frac{q_i^2 \, e^{q_i^2 t} \, \operatorname{erfc}(q_i \sqrt{t})}
{\displaystyle\prod_{j=1, j \ne i}^{4} (q_i - q_j)} \, ,
\end{equation}
and $q$, indexed by $i$ and $j$, runs over the four complex roots of the expression

\begin{equation}
    Ms^2 + \gamma\tau_fs^{3/2}+\gamma s+K \, .
\end{equation}

\subsubsection{History and Thermal Force statistical analysis}
As a consequence of the fluctuation-dissipation theorem, we have shown in the main text that for a free particle
\begin{equation}
    \langle R(t)v(0) \rangle = - \gamma\sqrt{\frac{\tau_f}{\pi}}\int_{-\infty}^0\frac{\langle v(0)v(t') \rangle }{2({t -t'})^{3/2}}dt' \, .
\end{equation}
By assuming that $R(t)$ and $v(t)$ together form a high-dimensional multivariate Gaussian space, this result is sufficient to determine the expected mean trajectories and MSDs for preconditioned values for $v(0)$. As a direct consequence of Gaussianity, it follows that
\begin{align}
\label{r_given_v0}
    \langle R(t) | v(0) \rangle = \langle R(t) v(0) \rangle \frac{v(0)}{\langle v^2(0)\rangle} = - \frac{v(0)}{\langle v^2(0)\rangle} \gamma\sqrt{\frac{\tau_f}{\pi}}\int_{-\infty}^0\frac{\langle v(0)v(t') \rangle }{2({t -t'})^{3/2}}dt' \\
    \langle I(t) | v(0) \rangle= \frac{v(0)}{\langle v^2(0)\rangle} \gamma\sqrt{\frac{\tau_f}{\pi}}\int_{-\infty}^0\frac{\langle v(0)v(t') \rangle }{2({t -t'})^{3/2}}dt' \, .
\end{align}
These results are sufficient to demonstrate that the mean trajectory is agnostic to the details of the history; the conditioned thermal force hides the details of this past history from showing up in mean trajectories. We can apply the same analysis for the particle's MSD, where now we use the fact that

\begin{align}
\label{matrix_conditioned}
    \langle R(t_1)R(t_2)|v(0) \rangle = \langle R(t_1) R(t_2) \rangle+ \langle R(t_1)|v(0) \rangle\langle R(t_2) | v(0)\rangle - \frac{\langle R(t_1) v(0) \rangle \langle R(t_2) v(0)\rangle}{\langle v^2 \rangle} \, . 
\end{align}


With these results for the conditional correlations, we can analyze the impact of the history on the MSD conditioned on the initial velocity of the particle. We will begin by analyzing the motion of a free particle with $x(0)=0$ for simplicity, and solve for $\langle x(t_1)x(t_2) |v(0)\rangle$. Then the full MSD including the contributions of the history and the conditional thermal force becomes

\begin{multline}
\label{msd_conditioned_1}
     \langle x(t_1)x(t_2) |v(0)\rangle = M^2v(0)^2B_f(t_1)B_f(t_2) + Mv(0)B_f(t_1)\int_0^{t_2}B_f(t_2-t')\langle R(t')|v(0)\rangle dt' \\
     + Mv(0)B_f(t_1)\int_0^{t_2}B_f(t_2-t')\langle I(t')|v(0)\rangle dt'+ Mv(0)B_f(t_2)\int_0^{t_1}B_f(t_1-t')\langle R(t')|v(0)\rangle dt'\\
     +Mv(0)B_f(t_2)\int_0^{t_1}B_f(t_1-t')\langle I(t')|v(0)\rangle dt'+\int_0^{t_1}B_f(t_1-t')\int_0^{t_2}B_f(t_2-t'')\langle R(t')R(t'')|v(0)\rangle dt''dt' \\
     +\int_0^{t_1}B_f(t_1-t')\int_0^{t_2}B_f(t_2-t'')\langle R(t')I(t'')|v(0)\rangle dt''dt'\\
     +\int_0^{t_1}B_f(t_1-t')\int_0^{t_2}B_f(t_2-t'')\langle I(t')R(t'')|v(0)\rangle dt''dt'\\
     +\int_0^{t_1}B_f(t_1-t')\int_0^{t_2}B_f(t_2-t'')\langle I(t')I(t'')|v(0)\rangle dt''dt' \, ,
\end{multline}
where $B_f$ is the Green's function for the free Brownian particle. The general form for a correlation function conditioned on the initial velocity is

\begin{equation}
\label{conditional_general}
    \langle Y(t_1)Z(t_2)|v(0) \rangle = \langle Y(t_1)Z(t_2) \rangle + \frac{\langle Y(t_1)v(0)\rangle \langle Z(t_2)v(0)\rangle}{\langle v(0)^2\rangle}(\frac{v(0)^2}{\langle v(0)^2\rangle}-1) \, .
\end{equation}

Equation \ref{r_given_v0} and \ref{conditional_general} allow us to evaluate the conditional correlation terms in equation \ref{msd_conditioned_1}, simplifying the conditioned MSD to
\begin{multline}
\label{msd_conditioned_2}
     \langle x(t_1)x(t_2)|v(0) \rangle = M^2v(0)^2B_f(t_1)B_f(t_2) +
     \int_0^{t_1}B_f(t_1-t')\int_0^{t_2}B_f(t_2-t'')\langle R(t')R(t'')\rangle dt''dt'\\
     +\int_0^{t_1}B_f(t_1-t')\int_0^{t_2}B_f(t_2-t'')\langle R(t')I(t'')\rangle dt''dt'
     +\int_0^{t_1}B_f(t_1-t')\int_0^{t_2}B_f(t_2-t'')\langle I(t')R(t'')\rangle dt''dt'\\
     +\int_0^{t_1}B_f(t_1-t')\int_0^{t_2}B_f(t_2-t'')\langle I(t')I(t'')\rangle dt''dt' \, .
\end{multline}
The cancellation between terms results in dependence only on $v(0)$ and equilibrium correlations.

After using the definition of $I(t)$, we are left with a conditioned MSD involving the VACF, $C_{vv}(t)$, and the correlation $\langle R(t)v(\tau) \rangle$. By time translation invariance, this term is $\langle R(t-\tau)v(0) \rangle$. Using equation \ref{equil_RV_condition} we now have

\begin{multline}
\label{msd_conditioned_4}
     \langle x(t_1)x(t_2)|v(0) \rangle = M^2v(0)^2B_f(t_1)B_f(t_2) +
     \int_0^{t_1}B_f(t_1-t')\int_0^{t_2}B_f(t_2-t'')\langle R(t')R(t'')\rangle dt''dt'\\
     -\gamma^2\frac{\tau_f}{\pi}\int_0^{t_1}B_f(t_1-t')\int_0^{t_2}B_f(t_2-t'')\int_{-\infty}^{0}\int_{-\infty}^{0}\frac{C_{vv}(\rho)}{4(t'-\tau-\rho)^{3/2}(t''-\tau)^{3/2}} d\rho d\tau dt''dt'\\
     -\gamma^2\frac{\tau_f}{\pi}\int_0^{t_1}B_f(t_1-t')\int_0^{t_2}B_f(t_2-t'') \int_{-\infty}^{0}\int_{-\infty}^{0}\frac{C_{vv}(\rho)}{4(t''-\tau-\rho)^{3/2}(t'-\tau)^{3/2}} d\rho d\tau dt''dt'\\
     +\gamma^2\frac{\tau_f}{\pi}\int_0^{t_1}B_f(t_1-t')\int_0^{t_2}B_f(t_2-t'')\int_{-\infty}^{0}\int_{-\infty}^{0}\frac{\langle v(\rho)v(\tau)\rangle}{4(t'-\rho)^{3/2}(t''-\tau)^{3/2}} d\rho d\tau  dt''dt' \, .
\end{multline}
Consider the very last integral in this expression. We can rewrite it as
\begin{equation}
    2\gamma^2\frac{\tau_f}{\pi}\int_0^{t_1}B_f(t_1-t')\int_0^{t_2}B_f(t_2-t'')\int_{-\infty}^{0}\int_{-\infty}^{\tau}\frac{\langle v(\rho)v(\tau)\rangle}{4(t'-\rho)^{3/2}(t''-\tau)^{3/2}} d\rho d\tau  dt''dt' \, .
\end{equation}
Since in equilibrium the VACF is stationary, we can keep the value of the integral the same, as long as all the time differences are kept consistent, to achieve
\begin{equation}
        2\gamma^2\frac{\tau_f}{\pi}\int_0^{t_1}B_f(t_1-t')\int_0^{t_2}B_f(t_2-t'')\int_{-\infty}^{0}\int_{-\infty}^{0}\frac{C_{vv}(\rho)}{4(t'-\rho - \tau)^{3/2}(t''-\tau)^{3/2}} d\rho d\tau  dt''dt' \, .
\end{equation}
This integral exactly cancels the other two terms involving the past dynamics of the particle. Therefore, for a free particle, we get perfect cancellation between the leftover history term and the information inferred about the stochastic force from knowledge of the initial velocity.

\subsubsection{Inclusion of the trapping potential}

It is not obvious that this cancellation with the thermal force works equivalently when the confining optical tweezer is added to the hydrodynamic Langevin equation. We repeat the analysis from the main text with the addition of the harmonic trap. The full Langevin equation becomes
\begin{equation}
    M\dot{v}(t) = -\gamma v(t) - \gamma \sqrt{\frac{\tau_f}{\pi}}\frac{v(0)}{\sqrt{t}}  +\gamma \sqrt{\frac{\tau_f}{\pi}} \int_{-\infty}^0\frac{v(t')}{2({t -t'})^{3/2}}dt' -\gamma \sqrt{\frac{\tau_f}{\pi}} \int_0^t\frac{\dot{v}(t')}{\sqrt{t -t'}}dt'+ R(t) - Kx(t) \, .
\end{equation}

Repeating the analysis from the main text, we multiply by $v(0)$ and take the Laplace transform to get
\begin{equation}
    \tilde{C}_{vv}(s)\big[Ms +\gamma +\gamma\sqrt{\tau_fs} + \frac{K}{s}\big] = MC_{vv}(0)  - K\frac{\langle x(0) v(0) \rangle }{s} +\mathscr{L}[\langle R(t)v(0) \rangle + \gamma\sqrt{\frac{\tau_f}{\pi}}\int_{-\infty}^0\frac{\langle v(0)v(t') \rangle }{2({t -t'})^{3/2}}dt'] \, .
\end{equation}
Under the equilibrium assumption, we expect the cross correlation $\langle x(0) v(0) \rangle$ to be zero so that
\begin{equation}
    \tilde{C}_{vv}(s) = \tilde{\mu}_K(s)\big[k_BT +\mathscr{L}[\langle R(t)v(0) \rangle + \gamma\sqrt{\frac{\tau_f}{\pi}}\int_{-\infty}^0\frac{\langle v(0)v(t') \rangle }{2({t -t'})^{3/2}}dt']\big] \, ,
\end{equation}
where $\tilde{\mu}_K(s)$ is the admittance for the trapped Brownian particle. The fluctuation-dissipation theorem then gives the identical result for the cross correlation $\langle R(t) v(0) \rangle$ as the case of a free particle. Since $\langle x(0) v(0) \rangle = 0$, the fact that $\langle R(t) v(0) \rangle$ takes the same value is sufficient to show that the same cancellation occurs for the trapped particle as the free particle. Thus, when calculating the MSD, terms involving correlations between the history, itself, the thermal force, and the velocity must all cancel out in the exact same way as above. Furthermore, we can still treat $\langle R(t) R(t')\rangle$ to have its equilibrium correlation properties. 

To find the effect of the trapping potential on the MSD, we return to equation \ref{x_solution}. Since we only condition on the velocity, we expect $\langle x^2(0) \rangle = \frac{k_BT}{K}$. The resulting expression for the conditioned position autocorrelation function becomes

\begin{multline}
\label{final_conditioned_msd}
    \langle x(t_1)x(t_2)|v(0)\rangle = k_BT[C(t_1)+C(t_2)-C(|t_2-t_1|)-MB(t_1)B(t_2)-KC(t_1)C(t_2)] + M^2v(0)^2B(t_1)B(t_2) \\
    + \frac{k_BT}{K}\big[M^2A(t_1)A(t_2) +\gamma^2 B(t_1)B(t_2) +\gamma^2\tau_f D(t_1)D(t_2) + M\gamma [A(t_1)B(t_2)+B(t_1)A(t_2)] \\
    + M\gamma \sqrt{\tau_f}[A(t_1)D(t_2)+D(t_1)A(t_2)] + \gamma^2 \sqrt{\tau_f}[B(t_1)D(t_2)+D(t_1)B(t_2)]\big] + \mathcal{F}(t_1, t_2) \, ,
\end{multline}
where $C(t)$ is defined by 
\begin{equation}
C(t) = \frac{1}{K}+\frac{1}{M} \sum_{i=1}^{4} 
\frac{e^{q_i^2 t} \, \operatorname{erfc}(q_i \sqrt{t})}
{q_i\displaystyle\prod_{j=1, j \ne i}^{4} (q_i - q_j)} \, ,
\end{equation}
and the final term, $\mathcal{F}(t_1, t_2)$, is given as

\begin{multline}
    \mathcal{F}(t_1, t_2) = \langle [Mx(0)A(t_1)+\gamma x(0)B(t_1) + \gamma \sqrt{\tau_f}x(0)D(t_1)][\int_{0}^{t_2}B(t_2-t')I(t')dt' +\int_{0}^{t_2}B(t_2-t')R(t')dt'] \\
    + [Mx(0)A(t_2)+\gamma x(0)B(t_2) + \gamma \sqrt{\tau_f}x(0)D(t_2)][\int_{0}^{t_1}B(t_1-t')I(t')dt' +\int_{0}^{t_1}B(t_1-t')R(t')dt'] \rangle \, .
\end{multline}

Note that $\mathcal{F}(t_1, t_2)$ contains twelve cross terms involving the equilibrium correlation between the initial position and either the thermal force or the history force. Due to the separation in time scales between the trap dynamics and the velocity dissipation, we expect these terms' contribution to be small. We can calculate their contribution by following a procedure similar to the case of the velocity degree of freedom. As a consequence of linear response theory, we have that for the position autocorrelation function, $C_{xx}(t)$, the expression
\begin{equation}
    -\frac{d^2}{dt^2}C_{xx}(t) = \langle v(t)v(0)\rangle = k_BT \mu_K(t) 
\end{equation}
holds in equilibrium. Transforming this expression into the Laplace domain yields
\begin{equation}
    -s^2\tilde{C}_{xx}(s) + s C_{xx}(0) + \dot{C}_{xx}(0)  = k_BT \tilde{\mu}_K(s) .
\end{equation}

By using $C_{xx}(0) = \frac{k_B T}{K}$ and $\dot{C}_{xx}(0) = 0$, this provides an expression for the equilibrium position autocorrelation in the Laplace domain of
\begin{equation}
     \tilde{C}_{xx}(s) = \frac{k_B T}{K}\frac{Ms +\gamma +\gamma\sqrt{\tau_f s}}{Ms^2 +\gamma s + \gamma\sqrt{\tau s^3}+K}.
\end{equation}
Analogously to our argument for the thermal force-velocity cross correlation, we take our equation of motion, multiply both sides by $x(0)$, take the Laplace transform, and then take an ensemble average over equilibrium conditions. The resulting expression yields
\begin{multline}
    \tilde{C}_{xx}(s) =  \frac{k_B T}{K}\frac{Ms +\gamma +\gamma\sqrt{\tau_f s}}{Ms^2 +\gamma s + \gamma\sqrt{\tau s^3}+K}\\ +  \frac{1}{Ms^2 +\gamma s + \gamma\sqrt{\tau s^3}+K}\mathscr{L}[\langle R(t)x(0) \rangle + \gamma\sqrt{\frac{\tau_f}{\pi}}\int_{-\infty}^0\frac{\langle v(t')x(0) \rangle }{2({t -t'})^{3/2}}dt'].
\end{multline}
Utilizing our expression for the position autocorrelation function immediately yields the result quoted in the main text of 
\begin{equation}
    \langle R(t)x(0) \rangle = -\gamma\sqrt{\frac{\tau_f}{\pi}}\int_{-\infty}^0\frac{\langle v(t')x(0) \rangle }{2({t -t'})^{3/2}}dt'
\end{equation}.

The above expression immediately implies that $\mathcal{F}(t_1, t_2)$ is 0. Therefore, the equation used for the calculation of the theoretical curves for the velocity-conditioned MSD is arrived at using equation \ref{final_conditioned_msd}, neglecting $\mathcal{F}(t_1, t_2)$, and the definition of the MSD,

\begin{equation}
    \operatorname{MSD}(t) = \langle [x(0)-x(t)]^2 \rangle \, .
\end{equation}

\subsubsection{White Noise Langevin Equation}
In the analysis of crossing times $t_c$, we compare the dynamics of the incompressible fluid with the Brownian motion of an equivalent particle in a forcing environment of white noise. In order to make this comparison, we require an analytic solution to this white-noise model system as well. The white-noise Langevin equation is defined as shown in the main text by
\begin{align}
 &m\ddot{x}(t) = -\gamma \dot{x}(t) - K x(t) + R(t)\\
&\gamma = 6\pi a \eta \\
&\langle R(t)R(t') \rangle = 2k_BT\gamma\delta(t - t') \, .
\end{align}
This stochastic differential equation can be solved exactly for an arbitrary initial velocity and position. Doing so in the case of an overdamped trap yields the solution
\begin{multline}
     x(t) = e^{-t/2\tau}(x(0)\cosh(\omega_1 t) + \frac{1}{\omega_1}(v(0) + \frac{1}{2\tau}x(0))\sinh(\omega_1 t) \\+ \frac{\sqrt{2\gamma k_B T}}{m \omega_1}\int_0^t\sinh(\omega_1 (t-s))e^{s/2\tau}dW_s) \, ,
 \end{multline}
 where $\omega_1 \equiv \sqrt{-\omega_0^2 + \frac{1}{4\tau^2}}$, and $dW_s$ is the increment of a Wiener process. From here, one can calculate the MSD for arbitrary initial conditions. We are interested in the case of a set initial velocity and thermally distributed initial position. For these initial conditions, the MSD is set by three terms associated with the initial spread of the position, the initial velocity, and the influence of the thermal force. The sum of these three effects then yields the full MSD, which takes the form
\begin{multline}
    \operatorname{MSD}[t] = \frac{2  k_B T e^{-t/\tau}}{m\omega_1^2} \frac{-1 -4\tau^2\omega_1^2(-1 + e^{t/\tau}) + \cosh(2t\omega_1) + 2\tau \omega_1 \sinh(2 t \omega_1)}{8 \tau^2\omega_1^2 -2} \\+ \langle x(0)^2 \rangle [1 - 2e^{-t/{2\tau}}( \cosh (\omega_1 t) + \frac{1}{2 \tau \omega_1} \sinh ( \omega_1 t)) + e^{-t/\tau}( \cosh^2 ( \omega_1 t)\\ + \frac{1}{4 \tau^2 \omega_1^2} \sinh^2(\omega_1 t) + \frac{1}{\omega_1 \tau} \sinh(\omega_1 t) \cosh(\omega_1 t))] +  v(0)^2  \frac{1}{\omega_1^2} \sinh ( \omega_1 t)^2 e^{-t/\tau} \, .
\end{multline}

\subsubsection{Uncertainty Estimates for Correlated Trajectories}
When forming our ensembles of trajectories for a given initial velocity, nearby trajectories are necessarily correlated. Therefore, we cannot assume independence for all $N$ of our experimental trajectories when calculating the uncertainty of our measurements. Usually, these correlations are handled with the blocking method, as we have done when calculating our equilibrium MSD \cite{flyvbjerg_error_1989}. However, since our trajectories are not pulled at regular intervals from the data set, it is not possible to directly implement this algorithm with the conditioned velocity ensembles. Instead, we note that if the sampled trajectories are far enough apart, the correlation between the trajectories decays away so that they are approximately independent. For short times, the dynamics are a consequence of the velocity degree of freedom, and therefore this correlation time is set by $\tau_p$. At longer times, the dynamics are set by the interaction with the trapping potential and therefore correlations are primarily due to the position degree of freedom. The position correlation time is set by the trap strength and the damping acting on the fluid so that $\tau_K = \gamma / K$. To estimate the number of trajectories sufficiently separated to be considered statistically independent, we divide our time traces up into bins with width equal to $\tau_K$. We then count how many bins contain the starting point for one of our velocity conditioned trajectories and use this to define the effective number of statistically independent trials for the experiment ($N_{eff}$). Our uncertainty estimate is given by
\begin{equation}
    s_v = \frac{\sigma_v}{\sqrt{N_{eff}}}
\end{equation}
where $s_v$ is our experimental uncertainty and $\sigma_v$ is the experimental standard deviation across all of the trajectories for a given initial velocity. Note that for short lag times, this is a very conservative estimate, as the velocity de-correlates much faster than the position. For the longest lag times, it is also a conservative estimate, since it treats trajectories beginning in the same bin as perfectly correlated. We find that for all of the velocity conditioned MSD, mean trajectory, and variation around the mean trajectory graphs, $N_{eff}/N > 0.32$, so the effect on the uncertainty estimate is never greater than a factor of $1.8$ times the uncertainty would be if we assumed all the traces were completely independent. Therefore, this conservative accounting of the correlation between trajectories has a limited effect on the experimental uncertainties. 






\clearpage 



\end{document}